# DEFECT STRUCTURE OF THE HIGH-DIELECTRIC-CONSTANT PEROVSKITE $CaCu_3Ti_4O_{12}$

L. Wu, Y. Zhu*, S. Park, S. Shapiro, and G. Shirane,

Brookhaven National Laboratory, Upton, NY 11973

J. Tafto

University Oslo, Blindern, 0316 Oslo, Norway

## Abstract

Using transmission electron microscopy (TEM) we studied $CaCu_3Ti_4O_{12}$, an intriguing material that exhibits a huge dielectric response, up to kilohertz frequencies, over a wide range of temperature. Neither in single crystals, nor in polycrystalline samples, including sintered bulk- and thin-films, did we observe the twin domains suggested in the literature. Nevertheless, in the single crystals, we saw a very high density of dislocations with a Burger vector of [110], as well as regions with cation disorder and planar defects with a displacement vector ¼[110]. In the polycrystalline samples, we observed many grain boundaries with oxygen deficiency, in comparison with the grain interior. The defect-related structural disorders and inhomogeneity, serving as an internal barrier layer capacitance (IBLC) in a semiconducting matrix, might explain the very large dielectric response of the material. Our TEM study of the structure defects in $CaCu_3Ti_4O_{12}$ supports a recently proposed morphological model with percolating conducting regions and blocking regions.

___________________________________

* corresponding author, email: zhu@bnl.gov

# 1. Introduction

In recent years, CaCu$_3$Ti$_4$O$_{12}$, or CCTO, has attracted much attention due to its extremely high dielectric constant, or permittivity, at low frequencies that persists virtually unchanged over a wide temperature range, from 100 to 600 K [1]. Different superlatives, like giant, colossal and enormous, were used in reporting this high dielectric response. Practically all investigators observed a relative dielectric constant, close to $10^4$, from static fields up to kilohertz frequencies. The high dielectric constant appears to be an inherent property of the material because it is barely influenced by the processing route used to synthesize single crystals or polycrystalline sintered powder, or thin films [1-4]. But now, after having scrutinized the material using x-ray-, optical-, and neutron diffraction techniques [1-5], impedance spectroscopy [6], and theoretical electronic structure calculations [7], most researchers find it unlikely that the large dielectric response is intrinsic. In this context, intrinsic means that the large response would be present in a perfectly stoichiometric, defect-free single-domain crystal of CCTO. Rather, they believe its origin is extrinsic and related to the 'real' microstructure, such as domain boundaries. In particular, the idea of twin boundaries was put forward based on the observation from single crystal x-ray diffraction [3] that pairs of reflections such as *(hkl)* and *(–khl)* reveal equal intensities. However, in a twin-free crystal, the intensities of such pairs should differ due to the lack of four-fold symmetry in CCTO with space group Im-3 (no. 204). Structure factor calculations suggest that the intensities of many such pairs, e.g., *(310)* and *(–130)* are very different. The idea that the high dielectric constant is caused by the presence of twin boundaries or other planar defects is supported by two completely different approaches: density functional theory (DFT) calculations of the electronic structure [7], and impedance spectroscopy measurements [6]. Independently they suggest a non-percolating semiconducting matrix caused by insulating layers across twins, anti-phase, or grain boundaries. The insulating layers are thought to cause internal boundary layer capacitance (IBLC) [6]. Recently, six different morphological models, in terms of percolating or nonpercolating conducting layers, were offered as possible candidates to explain the atypical dielectrical behavior [8]

In the simple IBLC model, grain boundaries cannot be the sole boundary because single crystals exhibit similar dielectric properties. However, except for grain boundaries in polycrystalline materials of CCTO, domain boundaries are not easily visible in this material using conventional transmission electron microscopy (TEM). Thus, the search for domain boundaries calls for another strategy. In this paper, we first describe the types of possible domain boundaries expected in CCTO based on its crystal symmetry. Then, using a variety of TEM techniques, we test for the presence of such domain boundaries, namely, twin boundaries, anti-phase boundaries, and their combinations. Finally, we present our observations of other types of defects including dislocations, cation disorder, and grain boundaries observed in CCTO.

**2. Crystal structure and possible domain boundaries**

The body-centered cubic unit cell of $CaCu_3Ti_4O_{12}$ has a lattice parameter of 0.7405 nm at room temperature (see [1] for the structure parameters used in this study) that are 2x2x2 times the unit cell of the perovskite sub-cell. Following the notation used for perovskites, Ca and Cu are at A sites and Ti at B sites. The Ti atoms are octahedrally coordinated with the octahedra rotated the same amount around the [100], [010], and [001] axes. According to Glazer's notation [9], the tilt is $a^+a^+a^+$. The tilt is fairly large causing 3/4 of the A atoms, i.e., the Cu atoms, to be four-coordinated with the four oxygen atoms forming a square with a Cu atom at its center. The normal to the square is the [100] direction, or the equivalent [010] or [001] directions. The distance from the Cu atom to each of the four surrounding oxygen atoms is 0.198 nm. The remaining A atoms, i.e., the Ca atoms have a bcc arrangement, and each is surrounded by 12 oxygen atoms at a distance of 0.261 nm. At the bottom left of Fig. 1, we show the structure of CCTO projected along the [001] direction as indicated by the tilted $TiO_6$ octahedra and $CuO_4$ squares. The $CuO_4$ squares with centers at z=0 and z=1/2 are represented by the full and dotted lines, respectively, and normal to the sheet of paper show up as double-lines. Note the two-fold symmetry of the projected structure, showing that this cubic structure is not invariant to a rotation of $90°$. It is the arrangement of the oxygen atoms that breaks the four-fold symmetry. If domains related by $90°$ rotation are present within the crystal, then we are dealing with $90°$ rotation twins. In addition to such twins, conceivably there may be anti-phase domain boundaries in the structure with the unit cell displaced a vector of R=[1/2 0 0] or R=[1/2 1/2 0]. Fig.1 shows possible domain boundaries based on this crystal symmetry and orientation variant argument. There are five possibilities: the $90°$ rotation twin boundary; the anti-phase boundary associated with R=[1/2 0 0]; and with R=[1/2 1/2 0]; the combination of $90°$ rotation and R=[1/2 0 0]; and, $90°$ rotation combined with R=[1/2 1/2 0]. When these domains are present in sufficient density, i.e., not separated by more than a few tens of μm, they can be easily observed using a TEM. If the boundaries are scarcer, the likelihood of encountering them decreases in the TEM's limited field of view. Due to the crystal's cubic symmetry, the $90°$ rotation twin domains will not split the diffraction spots, regardless of whether or not they are combined with a translation displacement. Thus, it will be very difficult to observe such boundaries using conventional dark-field imaging that uses a well separated and distinct diffraction spot. Nevertheless, we note the structure factors of many reflections in the same-family of the cubic CCTO differ; exceptions are the *(hhl)* reflections. For example, for {310} type reflections, although the (-130) and (-310)$_T$ (subscript T refers to the (110) twin) reflections overlap in reciprocal space their structure factors are quite different ($F_{-130}$=17.23, $F_{-310T}$ = -10.18). The same applies to the {530} type reflections. This suggests that an imaging method based on the structure factor may reveal these twin domains and their boundaries. An anti-phase boundary gives contrast for reflections when **R·g** (where **R** is the displacement vector and **g** is the reflection used for imaging) differs from zero, or an integer. Again, there are numerous possible choices of **g**-vectors giving contrast.

3. Experimental

The TEM experiments were done with a JEOL 3000F electron microscope operated at 300kV, and equipped with a coherent electron source (field emission gun) and a high-resolution objective lens. Electron energy-loss spectroscopy (EELS) was performed using a Gatan energy-filter system. Diffraction simulations were carried out using both multi-slice and Bloch wave approaches based on computer codes developed at Brookhaven. Polycrystalline pellet sample from sintered powder, single crystal and thin film of CCTO were studied. The powder sample was prepared by heating a stoichiometric mixture of $CaCO_3$ (Analyzed Reagent 99.9%), CuO (Alfa 99.995%), and $TiO_2$ (Alfa 99.99%) at 1050 °C for 16 h with intermediate grinding at 800 °C for 1h. The sintered powder then was pressed into pellet. The CCTO thin film was prepared by Dr. Weidong Si, Physics Department, BNL, using pulse-laser deposition method (see [4]). The film was structurally inhomogeneous with large amount of as-grown defects and associated lattice distortion. The single crystal was grown by S. Wakimoto using traveling-solvent floating-zone method in an image furnace. The sample was previously used for the optical, neutron and x-ray measurements reported in [2]. For TEM observations, these samples were first cut into disks, then thinned and polished by ion-milling. Low-energy ion beam and plasma cleaner were used in the final stage of preparing the sample. The samples were studied at room temperature, and at 86K using a liquid-nitrogen cooling stage. Overall, the samples showed many similarities. However, the thin film had additional lattice defects associated with their expitaxial growth, while, in sintered materials, the size of the crystal grains ranged from submicrons to tens of microns. To avoid confusion from these as-grown defects in thin films, we discuss our observations from single crystals and sintered polycrystalline samples only. The advantage of studying the latter is that we have an easy access to crystal grains with various orientations.

## 4. Results
### 4.1 Search for orientation domains

Fig.2 (a) is the electron diffraction pattern from a small-angle convergent beam showing the two-fold symmetry along the [001] direction in agreement with the simulation (Fig.2b) based on the Im-3 space group for a crystal 9.2nm thick. Within the more than a dozen crystal grains we studied we did not encounter any 90$^o$ rotation of this pattern. Such a rotation would have signaled that we had crossed a 90$^o$ rotation twin boundary. Imaging with different reflections at the Bragg position in a bright-field and dark-field did not reveal any domains and domain boundaries. As mentioned early, because of its cubic symmetry, possible twinning in CCTO does not generate splitting of the Bragg spots; thus, the twins and the matrix are always imaged simultaneously. Even in relatively thin areas, multiple scattering can wash out the contrast difference between the two. A robust approach to image the twin domains in CCTO will be a method that can separate orientation domains and is sensitive to the structure factor. Fig.3 shows a simulated coherent shadow-image diffraction-pattern of a (110) twin boundary in CCTO using the PARODI method (parallel recording of dark-field images) we developed for mapping the distribution of valence electrons and interface analysis in superconductors [10,11]. In this method, the electron beam is focused above the sample thus illuminating an area of interest, including a defect such as an interface. We then record the images of many reflections simultaneously in the back-focal plane of the objective lens. The image intensities of these reflections strongly depend on their structure factors. When a coherent electron source is used strong interference fringes associated with the defect appear. In the calculation, as shown in Fig.3 in a (001)$^*$ zone pattern, we include an area of 26nm with the twin boundary in the middle. We observed remarkable interference fringes at the twin boundary in many diffraction disks where, across the boundary, the corresponding reflection with different structure factor changes, say from (-130) to (-310)$_T$ ($F_{-130}$=17.23, $F_{-310}$=-10.18). A similar situation applies to the {350} type of reflections. Since the structure factors for the (hh0) and (h00) type of reflections are the same, no twin-boundary contrast is visible.

The PARODI method is powerful in searching for twining that does not lead to the splitting of diffraction peaks, which is often required for conventional dark-field imaging in TEM. Nevertheless, after an extensive research using this method in thin areas of both single crystals and polycrystalline samples as well as thin films, we did not observe any orientation domain and domain boundaries.

**4.2 Dislocations**

We encountered dislocations in both sintered polycrystals and single crystals. The dislocation density is very low in the former, but rather high and varies with areas in the latter. Fig.4(a) shows the typical morphology of dislocations in a single crystal. They usually do not exhibit clear and sharp images, and often they are associated with complex strain contrast, which may be attributed to the material's mechanical properties. In some areas we also observed a tangled dislocation network. The dislocations show strong contrast with reflections of **g**=020, 200, or 101. To determine the Burgers vector **b** of the dislocations, we tilted the sample to various orientations to form a two-beam imaging condition using the extinction criterion **g·b**=0. Fig.4(b-d) shows an example, imaged for the same two dislocations using **g**= 020 (a), **g**=-101 (b) and **g**=-110 (c), respectively. We note that when **g**=-110 is used the dislocations are out of contrast, consistent with the Burgers vector of **b**=110. We characterized various dislocations in single crystals as well as in the sintered polycrystals and found the majority are the <110> type with a mixed edge-and-screw component. Although we did not determine the amplitude of the Burgers vector, based on the crystal symmetry it can be either │1/2[110]│, or │1/4[110]│, corresponding a lattice displacement between the A atoms (Cu and Ca), or between the A (Cu/Ca) atoms and B (Ti) atoms, respectively.

**4.3 Cation-disorder-induced planar defects in single crystal**

Although we did not observe the orientation domain boundaries proposed in Fig. 1, we did see planar-like defects and regions with significant cation disorder, especially the disorder of Ca and Cu atoms in single crystal CCTO. For a perfect CCTO, the Wyckoff position for Ca is 000, while for Cu is 1/2,1/2,0. The intensity of the {110} reflections (structure factor squared) can be determined to be 11.21. However, when the Ca and Cu atoms change from the ordered to the disordered state, that is, the two sites are randomly occupied by Ca or Cu, the intensity of {110} reduces drastically to 0.32, which becomes invisible in electron diffraction pattern. This cation disorder lowers the crystal's symmetry. Large-scale disorder can be easily observed in selected-area-diffraction patterns, while small area disorder can be seen in lattice images. Fig.5 shows two such areas of disorder. In Fig.5(a), a HREM (high-resolution electron microscopy) image viewed in the (111) projection illustrates a disordered region with a change of local crystal symmetry. Fourier analysis of the image suggests that the lattice disorder occurs along the [-101] and [01-1] directions in the left area (only the (-110) spot present), while along the [-110] and [01-1] directions in the right area (only the (-101) spot present). Another area is shown in Fig.5 (b) where the disorder occurs in the [-101] and [-110] directions (only the (01-1) spot present). A displacement of half spacing of the (110) lattice is clearly visible, as marked by two white lines. The displacement is easy to see when the lattices are viewed at a glancing angle. In Fig.5(c) an experimental selected-area-diffraction pattern from a much large area also shows the loss of the 3-fold symmetry (with the $ATiO_3$ sub-cell symmetry in two directions), as compared with a typical non-disordered CCTO diffraction pattern (Fig.5(d)). Since it is a local disorder, volume averaged x-ray diffraction cannot reveal such a disorder, but will instead show a false 3-fold symmetry.

In the [001] orientation, such a ½(110) lattice displacement caused by the disorder often takes the form of a planar defect as shown in a low magnification image (Fig 6(a)). An enlarged area is shown in Fig.6(b), with the (110) lattice viewed edge-on. Both sides of the lattice abutting to the defect have the same crystallographic orientations while the (110) lattices exhibit a half spacing shift (i.e., a ¼[110] lattice translation) when they cross it because they are scattered out-of-phase. We note that the interpretation of high-resolution images often is not straightforward and care must be taken in reaching conclusions. We also note that such lattice displacements exist in areas with no significant difference in thickness; it is unlikely they originate from image artifacts. An unusual temperature dependence of local atomic displacements of Ca and Cu in CCTO has been observed using pair distribution function analysis [5]. The possible origin of the displacements can be local cation disorder, i.e., exchange of A (Cu/Ca) and B (Ti) sites, as supported by our diffraction study. A structural model for such a planar defect is depicted in Fig.6(c) with the $TiO_6$ octahedra across the interface shifted ¼[110] due to the movement of Ti to the Ca/Cu site, as marked by the solid lines across the interface. To have the octahedral match at the interface, an additional ~6% lattice shift (d=0.063[110]) was introduced in the direction perpendicular to the boundary. The observed planar defect associated with the cation disorder reported here could have significant implications since the disorder-induced lattice discontinuity or displacement will cause IBLC and change the local dielectric properties of the material. Along with the high-density of dislocations that have same displacement direction of the planar defects, we believe these intrinsic structural defects are responsible for the large dielectric response observed in the single crystal CCTO.

**4.4 Grain boundaries in sintered samples**

The origin of the dielectric behavior in polycrystalline samples can be totally different from that in a single crystal since the polycrystals comprise many grain boundaries. The polycrystalline samples we studied have an averaged grains size of about 1μm, with few defects in the grain interior. To search for the composition and/or electronic structural difference between the grain interior and grain boundaries, we performed electron energy-loss spectroscopy (EELS) on these samples. The energy-loss spectra at 86 K from an area far away from the boundary are shown in Fig. 7, both from the low loss region, and from the core-loss (K-edge of oxygen and the L-edges of the metal atoms) regions. There was no significant difference from those obtained at room temperature. The low loss region contains information about the complex dielectric function of the material, covering the energy range from visible light to soft x-rays. The energy dependence of the real and imaginary part of the permittivity can be extracted from these spectra by Kramers-Kronig analysis [12]. The similarity of these spectra at 86 K and room temperature suggests that the temperature dependence of the dielectric function is limited to the frequencies lower than that can be resolved with our electron energy loss spectrometer. The near- edge structure of the K-edge of oxygen is not easily interpreted without electronic structure calculations in this oxide where three different metal atoms surround each oxygen atom. The white $L_{2,3}$ lines from the metal atoms are as expected. The typical crystal-field splitting effect is seen for the $L_{2,3}$ lines of Ti; each is split into two, which is typical for $Ti^{4+}$ at the center of an oxygen octahedron [13]. White lines from Cu depend strongly on the oxidation state. The features seen for the $L_{2,3}$ lines of Cu in this material are consistent with $Cu^{2+}$ [13].

In comparing the EELS spectra from grain boundary and grain interior, we focused on the oxygen concentration at the boundaries since the resistivity of an $ATiO_3$ perovskite is often extremely sensitive to its oxygen content. It was reported that the room temperature resistivity of $BaTiO_{3-\delta}$ ceramics decreases from about $10^{12}$ Ωcm for $\delta \sim 0$ to about 10 Ωcm for $\delta \leq 0.0002$ [6]. Under appropriate processing conditions, minimal reoxidation can produce thin insulating layers on the outer surfaces or along the grain boundaries of the material to form surface and internal barrier layer capacitors (IBLC) resulting in a high "effective" dielectric constant of more than $10^4$ [14].

Fig.8 shows a spectra pair for the Ti $L_{2,3}$- and O K-edge from sintered bulk pellets at room temperature. We used Ti concentration as a reference by comparing the O/Ti ratio after subtracting the background and correcting for multiple scattering of the spectra. By averaging the integrated intensity ratio (using a 50eV energy window at a 50 mrad scattering angle) for six boundaries, we found that the grain interior is O/Ti=3, while, for grain boundaries, the ratio ranges from 2.90 to 2.65. This suggests a 3-12% decrease in oxygen concentration at grain boundaries (the average is about 6% decrease). The observations imply that, in addition to the strain associated with misorientation at the grain boundaries and the possible off-stochiometric composition of cations, there is a considerable variation in the amount of oxygen at the boundaries, consistent with what was observed with the impedance spectroscopy measurements [6]. Since oxygen is known to be crucial to the local conductivity in peroveskite oxides, it is likely that the grains are conducting (or semiconducting) and the grain boundaries are insulating (or blocking), resulting in a high dielectric response. Nevertheless, although we cannot rule out that the reverse also is possible since conducting bulk and the conducting interface morphologies are equally plausible [8].

## 5. Discussions

The unusual dielectrical behavior displayed in CCTO has numerous potential technological applications; however, the origin of the high dielectric response is not fully understood. In contrast to what was expected from the literature [15], we did not observe twins or anti-phase domains either in the single crystal or in the polycrystalline CCTO using TEM. Based on the crystal symmetry argument, these orientation domains still may be there in single crystals, and we may not have encountered them using TEM if they are several hundreds of µm apart. Realistically, even such well separated domains do exist they probably are not responsible for the IBLC effect. In polycrystalline material, with typical grain size of 1 µm, the observed two-dimensional boundaries are mainly grain boundaries. Even if twin or anti-phase domains are present in a small fraction of the grains, they cannot prevent percolation.

The similar dielectric properties observed in single crystals and polycrystals was puzzling. Based on our microscopy observations, we now can attribute the cause of the barriers for electrical conductivity to different structural defects in the two different types of CCTO samples. For polycrystalline samples, the main defects are the grain boundaries, while for single crystals they are dislocations and planar defects caused by cation disorder. These defects can form IBLC and thus lead to the observed unusual dielectric properties. In discussing possible morphologies, Cohen et al. [8] point out: "It is important to recognize that the mechanism leading to large dielectric constants can differ in polycrystalline (ceramic) and single-crystal samples". These investigators propose six possible models for the large dielectric constants. Among them, the one with a non-percolating, conducting, or semiconducting bulk and non-conducting grain boundaries is consistent with our observations of the polycrystalline material where grain boundaries exhibit oxygen deficiency in comparison with the bulk. On the other hand, the high density of dislocations and the planar defects observed in our single crystal can result in lattice distortions that may be accompanied by heterogeneity in composition and dielectric properties. Thus, our observations on single crystals could fit their proposed model with percolating conducting regions and surface blocking. The breakdown of the IBLC at low temperature due to electronic localization associated with these defects may be attributed to the temperature dependence of the dielectric response in the material.

Finally, we would like to point out that recently high dielectric response and a drop by a factor of 100 of the dielectric response at low temperature has also been observed in cuprates and nickelates, and it was related to an "electronic glassy state" [16]. Charge inhomogenaties in complex oxides are due to competing interactions (i.e. spins, charges, and strain) and often results in non-uniform electronic, chemical and magnetic ground states. The unraveling of these defect structures using local probes such as electron diffraction and HREM is the basis for understanding of many intriguing material properties.

**Acknowledgement**

The authors would like to thank C.C. Homes and T. Vogt for their stimulating discussions and Weidong Si and S. Wakimoto for proving some of the samples studied. One of the authors, S. Park, would also like to thank T. Vogt for his encouragement. This work was supported by Division of Materials Sciences, Office of Basic Energy Science, U.S. Department of Energy, under contract No. DE-AC02-98CH10886.

**References**

1. M. A. Subramanian et al., J. Sol. State Chem. **151**, 323 (2000)
2. C. C. Homes et al., Science **293**,673 (2001)
3. M. A. Subramanian and A. W. Sleight, Sol. State Science **4**, 347 (2002)
4. W. Si et al, Appl. Phys. Lett. **81** 2056 (2002).
5. E. S. Bozin et al., J. Phys. : Condens. Matter **16** S5091 (2004)
6. D. C. Sinclair et al., Appl. Phys. Lett. **80**, 2153 (2002)
7. L. He et al. Phys. Rev. B **65**, 214112 (2002)
8. M. H. Cohen et al., J. Appl. Phys. **94**, 3299 (2003)
9. A.M. Glazer, Acta Crystallogr. B**28**, 3384 (1972)
10. L. Wu, Y. Zhu, and J. Tafto, Phys. Rev. Lett., **85**, 5126 (2000).
11. L. Wu, Y. Zhu, T. Vogt, H. Su, J.W. Davenport and J. Tafto, Phys. Rev. B**69** 064501 (2004).
12. R.F. Egerton, "Electron Energy-Loss Spectroscopy in the Electron Microscope", Plenum Press, New York, 1986
13. R. D. Leapman, L. A. Grunes and P. L. Fejes, Phys. Rev. B **26**, 614 (1982)
14. Y. Cheng-Fu, Jpn. J. Appl. Phys. Part 1, **35**, 1806 (1996), **36** 188 (1997).
15. C.C. Homes et al, Phys. Rev. B**67**, 092106 (2003).
16. T. Park et al cond-mat/0404446, April (2004).

**Figure captions**

Fig.1

Schematics of the five possible orientation domains in CCTO: the twin boundary (denoted as TB: II-IV), anti-phase boundary (denoted as APB: I-II, II-III), and combinations of twin and anti-phase boundaries (I-IV, III-IV) projected along the [001] direction. The squares represent the unit-cell with the diamond showing its two-fold symmetry. The detailed structure of the same projection with $TiO_6$ octahedra and $CuO_4$ squares is included at the bottom left. The $CuO_4$ square viewed end-on is shown as a double-line.

Fig.2

Experimental (a) and calculated (b) convergent beam electron diffraction pattern (convergent angle : 1.64 mrad) of $(001)^*$ zone showing a two-fold symmetry in the crystal.

Fig.3

A calculated coherent PARODI pattern showing the interference fringes of the (110) twin boundary and its adjacent twin domains in the $(001)^*$ zone orientation. Note, the vertical fringes originate from the twin boundary, and the circular fringes from the condenser aperture that defines the edge of the disks. The twin boundary, a 90°twinning (II-IV in Fig.1), is clearly visible in shadow images of the (310)-type reflections due to the difference in their structure factors ($F_{-310}=F_{-310T}=-10.18$, $F_{-130}=F_{-130T}=17.23$), while invisible in the (220)-type reflections ($F_{220}=F_{-220}=F_{-2-20T}=F_{-220T}=45.7$).

Fig.4

(a) A typical bright-field image showing a high density of dislocations observed in the single crystal. (b-d) diffraction contrast of the same two dislocations imagined using **g**=020 (b), **g**=-101 (c), and **g**=-110 (d) reflections. With **g**=-110, the dislocations are out of contrast.

Fig.5

(a) A high-resolution image viewed along the [111] direction in single crystal CCTO showing the loss of 3-fold symmetry, evident from the Fourier analysis of the image, with only one of the three (110) reflections present ((-110) in the left region and (-101) in the right). (b) Another area in the same viewing direction showing the presence of only the 01-1 reflection. The white lines indicate a lattice shift in the region. The displacement associated with the loss of symmetry is most visible when the images are viewed along a glancing angle. (c) Selected area diffraction from a much larger area (1μm across) showing the regional loss of the CCTO symmetry along the [01-1] direction.   The diffraction pattern from an undistorted CCTO in a relatively thin area is shown in (e) for reference.

Fig.6

(a) A planar defect observed in the single crystal. (b) The high-resolution image of the same defect in (a) showing a lattice shift (R=1/4[110]) adjacent to the defect.  (c) A structural mode in a [001] projection illustrates the possible origin of the planar defect with a lattice shift of R=1/4[110] caused by Ca/Cu cation disorder (between the A atoms) and the shift of the A an B (Ti atoms) site.  To fit the $TiO_6$ octahedra across the interface, a small displacement (d=0.0603[-110]) perpendicular to the interface has to be introduced.  The parallel lines indicate the ¼[110] lattice displacement.

Fig.7

Low temperature electron energy-loss spectra recorded at 86K from the polycrystalline sample. From left: low loss region, L-edge of Ca, L-edge of Ti, K-edge of O and L-edge of Cu.

Fig. 8

A pair of electron energy-loss spectra of the Ti $L_{2,3}$- and O K-edge acquired from a grain boundary and away from the boundary at room temperature after subtracting the background  and correcting for multiple scattering , showing a low O/Ti ratio at the grain boundary.

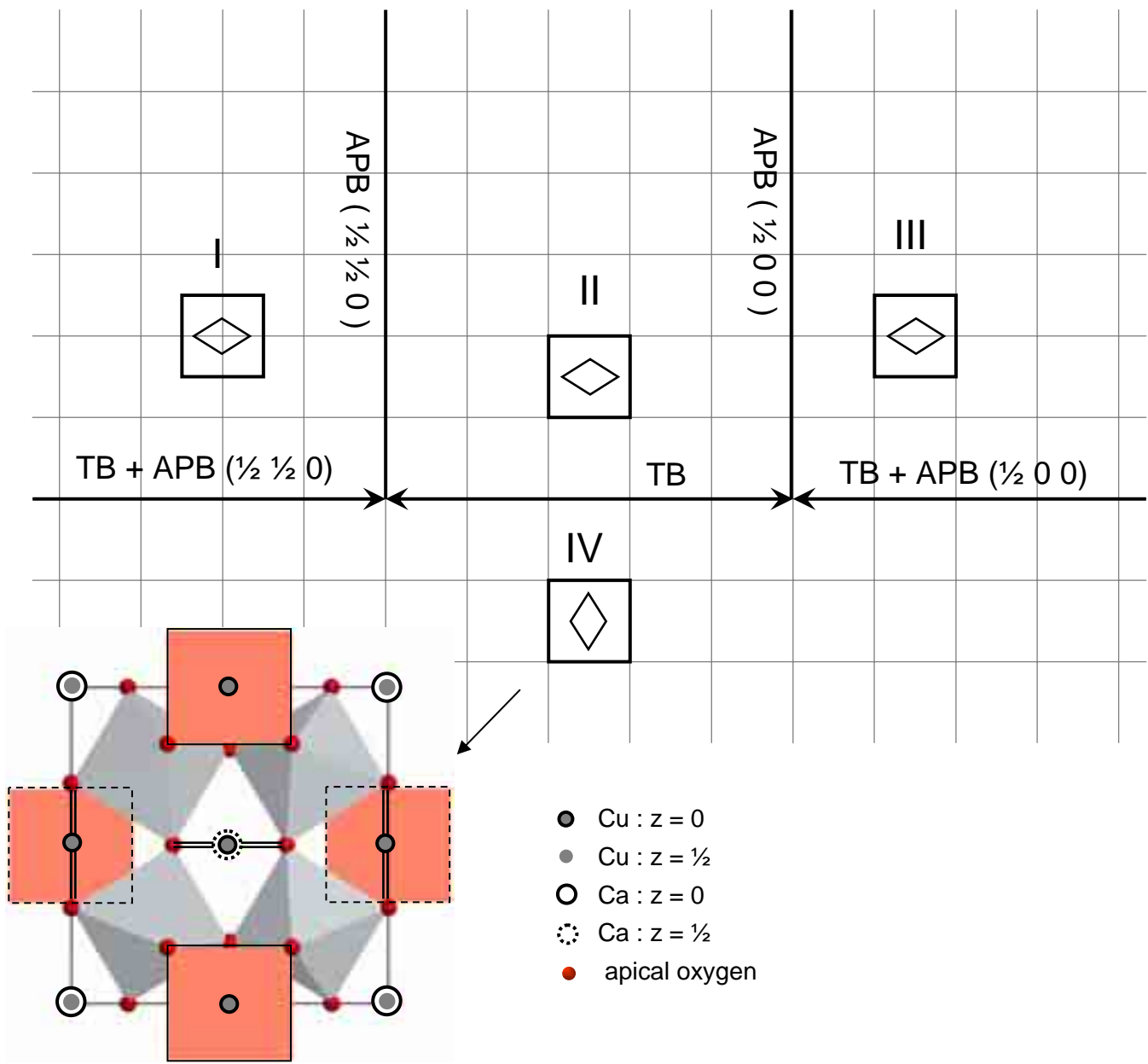

Fig.1

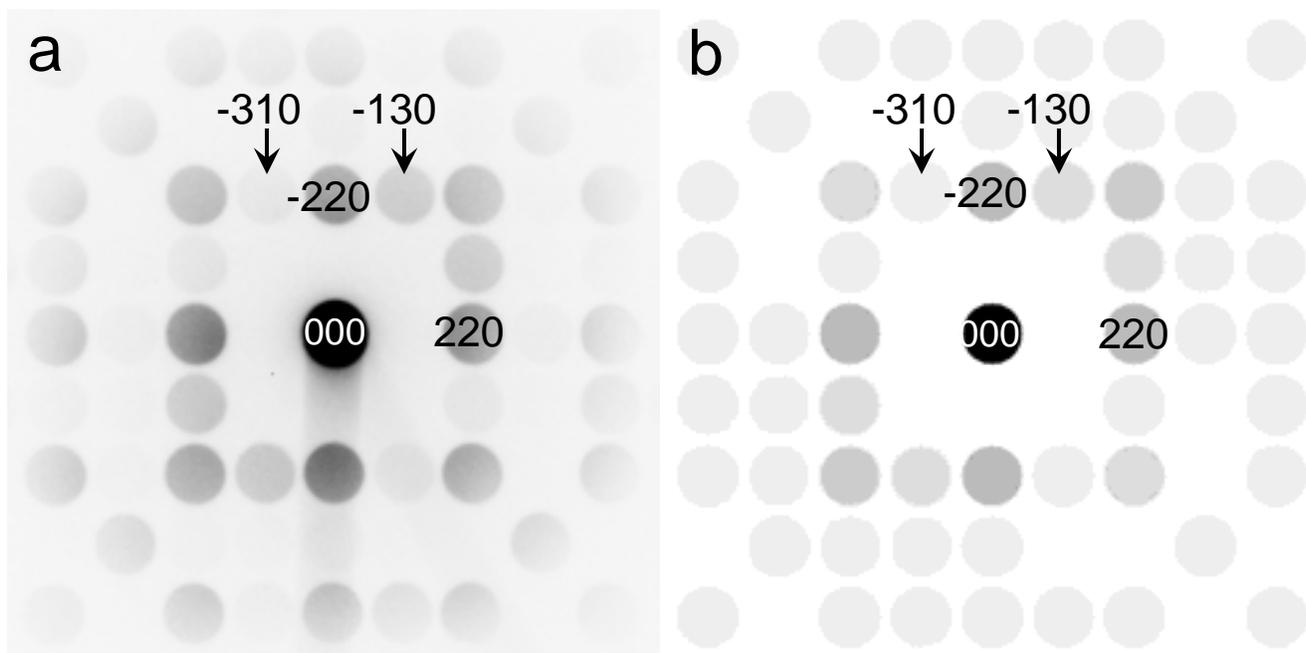

Fig.2

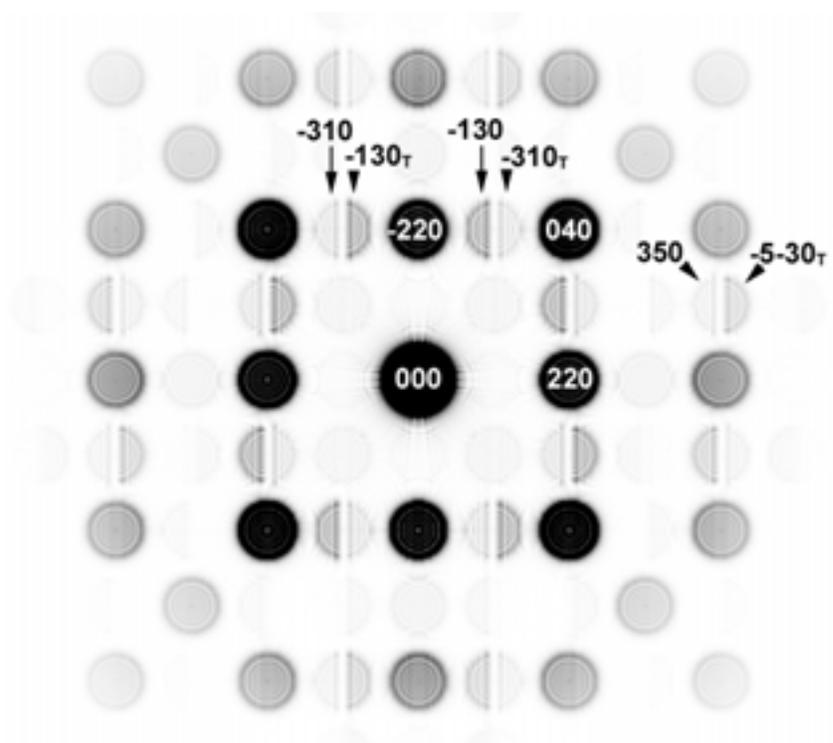

Fig.3

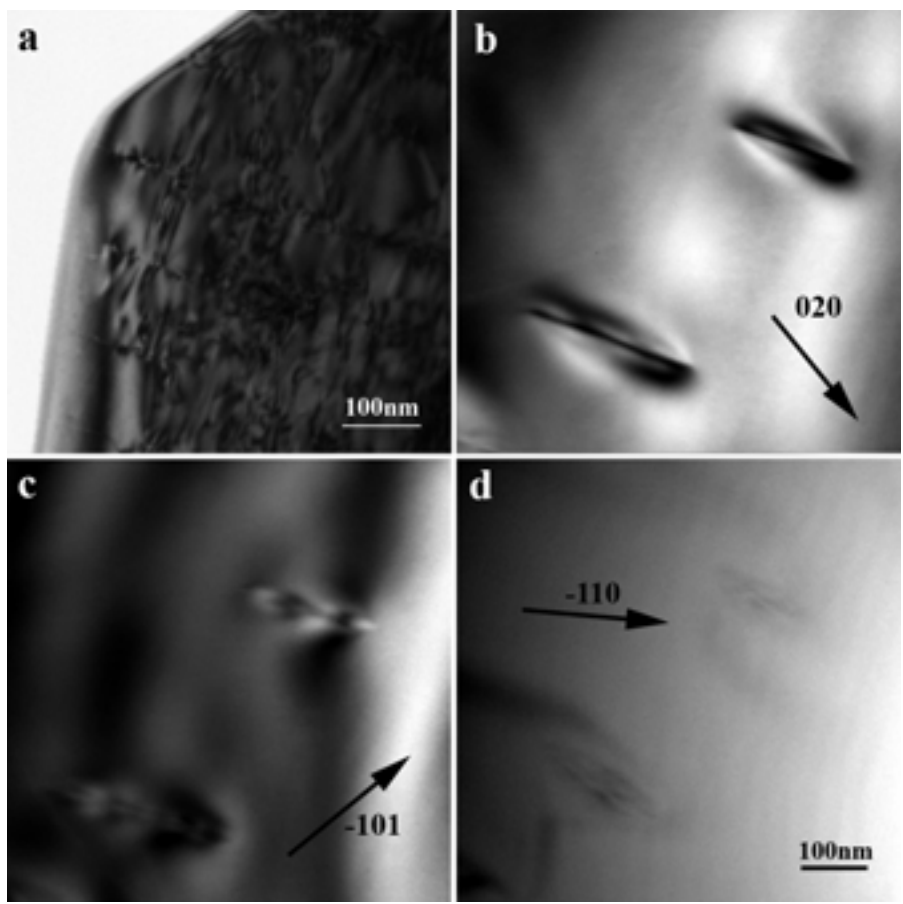

Fig.4

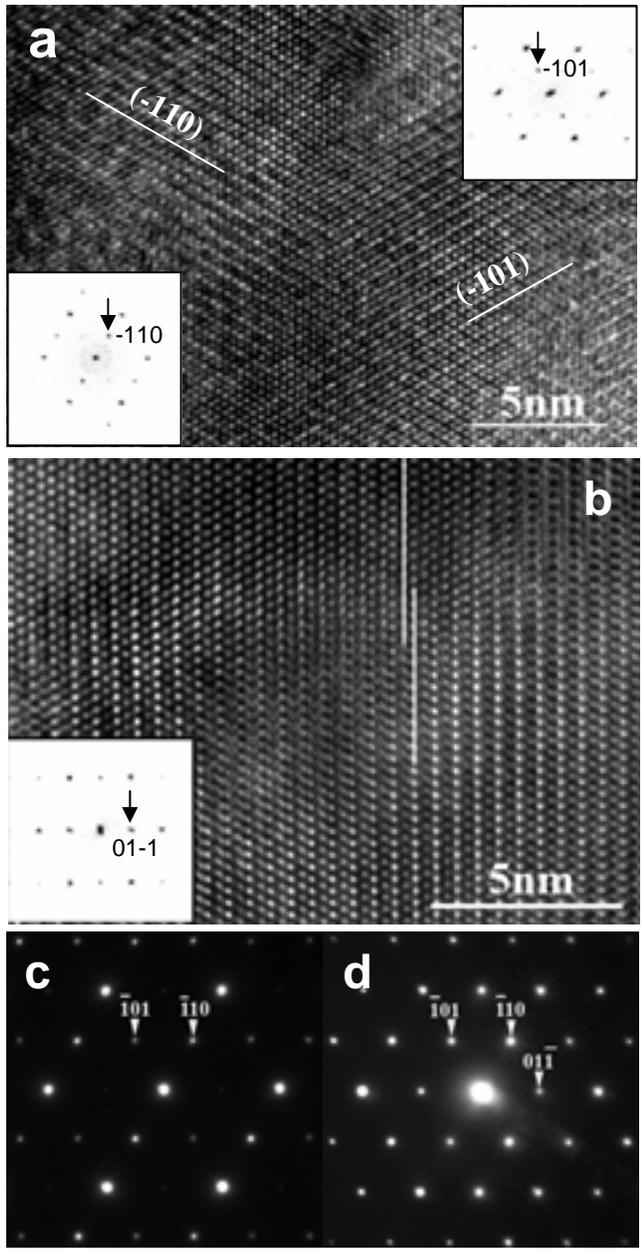

Fig.5

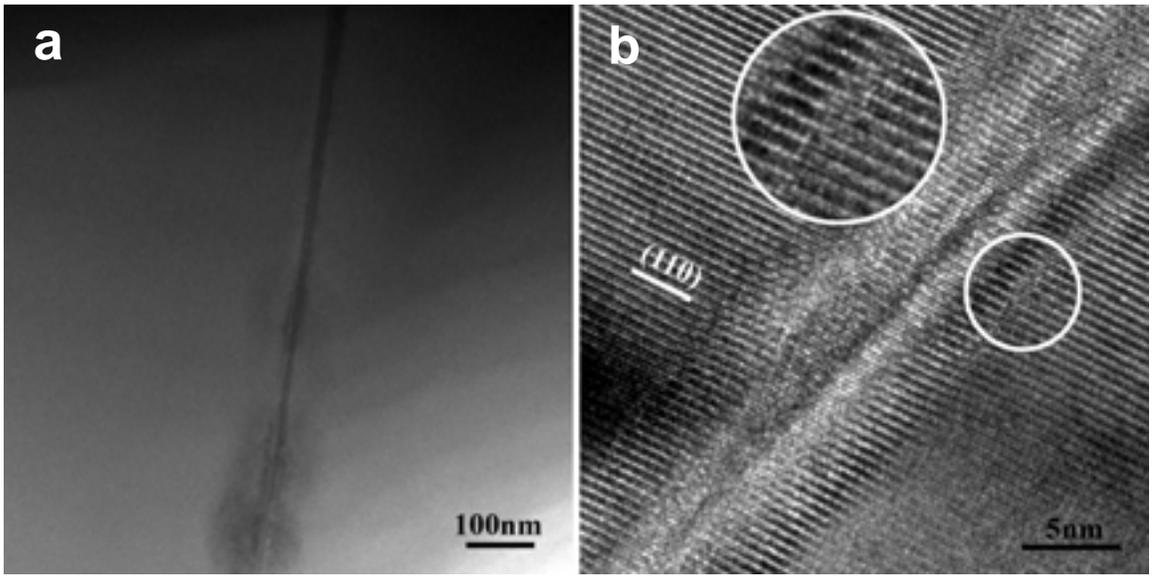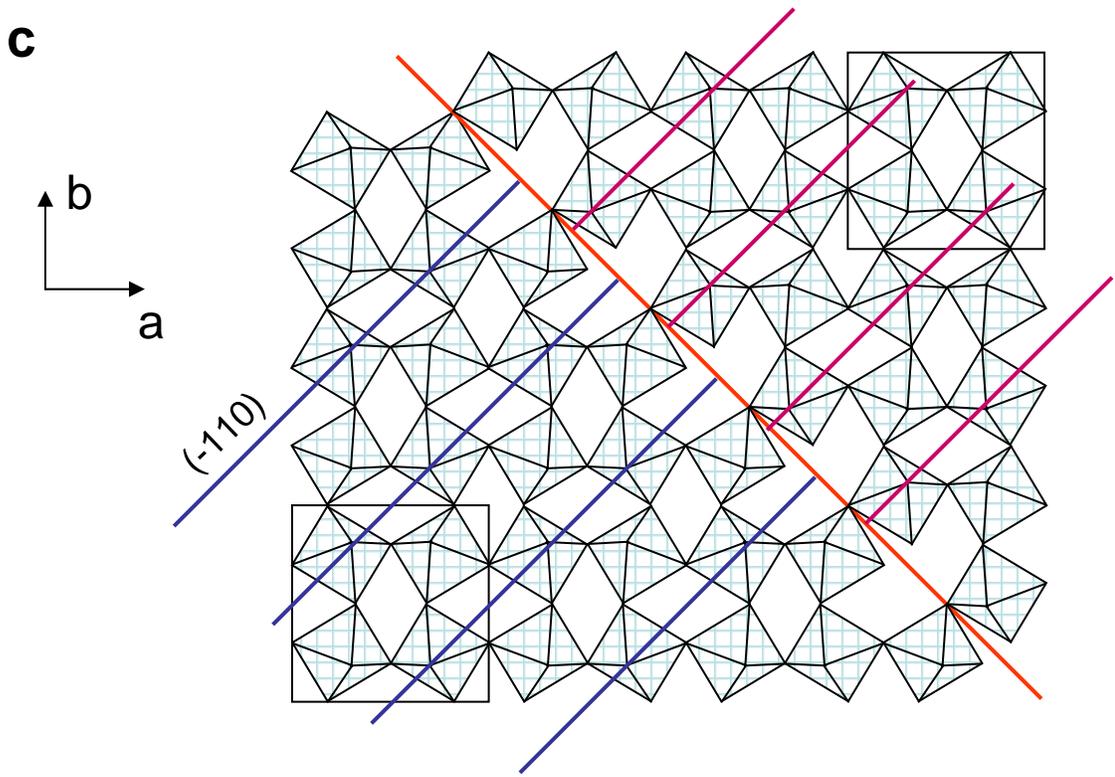

Fig.6

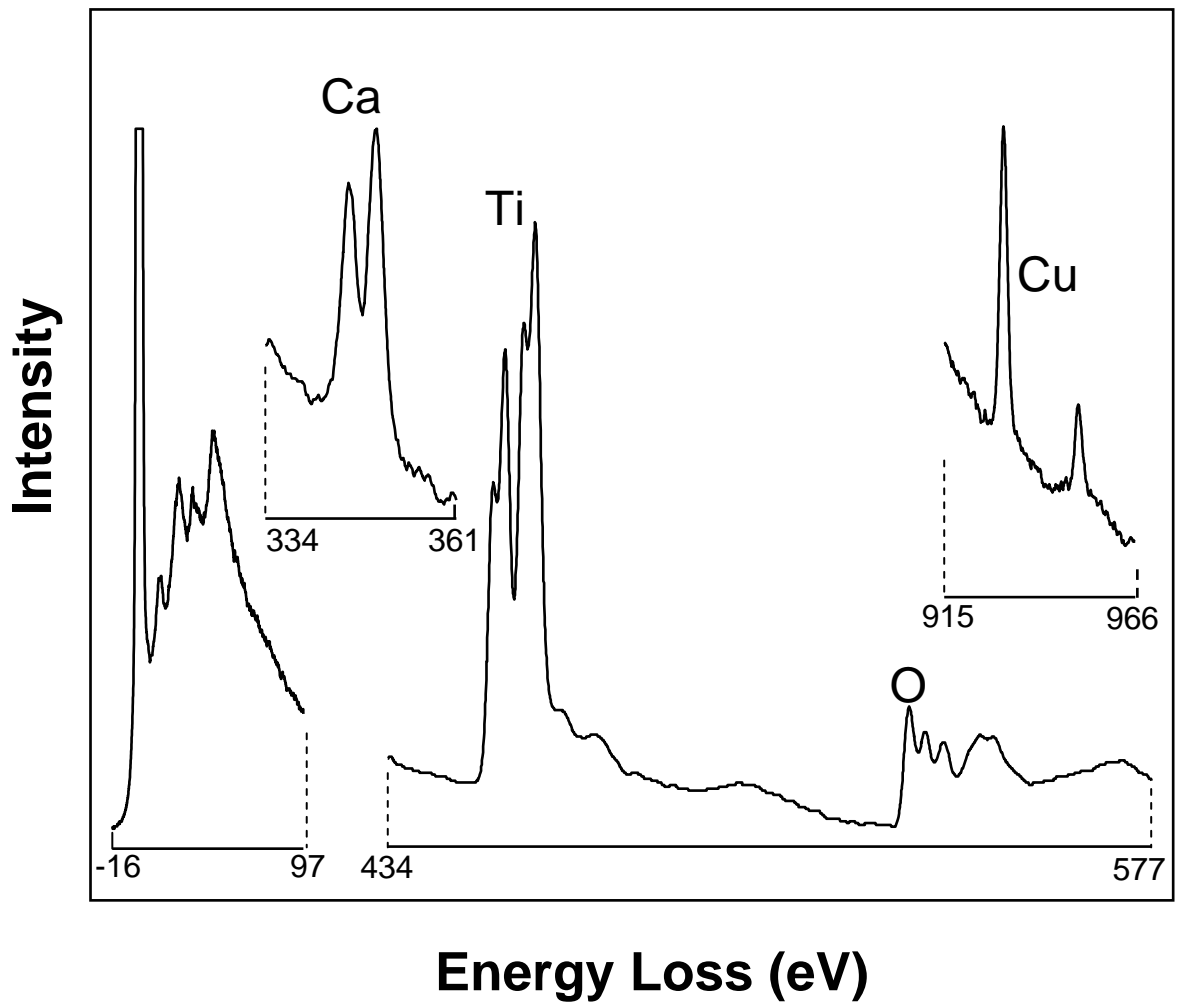

Fig.7

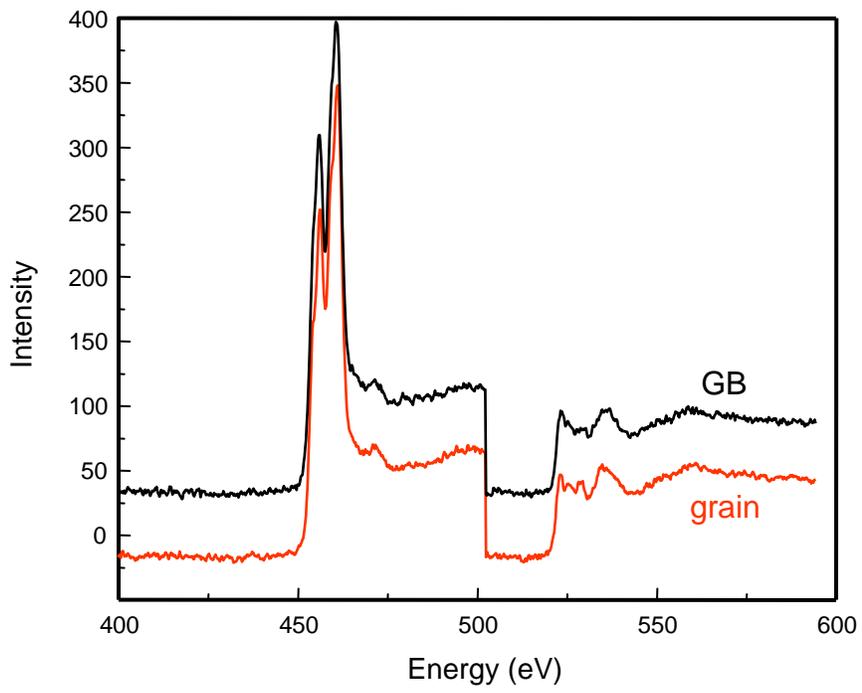

Fig.8